\documentclass[conference]{IEEEtran}
\IEEEoverridecommandlockouts
\usepackage{amsmath,amssymb,amsfonts}
\usepackage{algorithmic}
\usepackage{graphicx}
\usepackage{float}
\usepackage{textcomp}
\usepackage{xcolor}
\usepackage{balance}

\usepackage{fancyhdr}
\usepackage{graphicx,eso-pic}
\usepackage{lipsum}
\usepackage{wasysym}
\usepackage{marvosym}
\usepackage[alpine]{ifsym}

\AddToShipoutPictureBG{%
  \AtPageUpperLeft{%
    \raisebox{-\height}{\hspace{4.0em}\includegraphics[height=2.1cm]{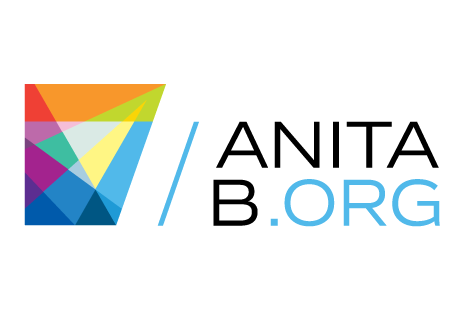}}
  }
 }

\begin{document}

\title{A Study of Few-Shot Audio Classification}

\author{
  \IEEEauthorblockN{
    Piper Wolters\IEEEauthorrefmark{3}\IEEEauthorrefmark{1}\thanks{\IEEEauthorrefmark{1}These authors contributed equally.}\thanks{This work was funded by the U.S. Government.}, 
    Chris Careaga\IEEEauthorrefmark{3}\IEEEauthorrefmark{1}, 
    Brian Hutchinson \IEEEauthorrefmark{3}\IEEEauthorrefmark{2} and
    Lauren Phillips\IEEEauthorrefmark{2}\\
      \IEEEauthorrefmark{3} Computer Science Department, Western Washington University, Bellingham, WA  \\
    \IEEEauthorrefmark{2}
      Pacific Northwest National Laboratory, Richland, WA
  }
}

\maketitle
\thispagestyle{fancy}
\cfoot{\small{\textbf{\textcolor[rgb]{0,.5,.69}{We envision a future where the people who imagine and build technology mirror the people and societies for whom they build it.}}}}

\begin{abstract}
Advances in deep learning have resulted in state-of-the-art performance for many audio classification tasks but, unlike humans, these systems traditionally require large amounts of data to make accurate predictions. Not every person or organization has access to those resources, and the organizations that do, like our field at large, do not reflect the demographics of our country. Enabling people to use machine learning without significant resource hurdles is important, because machine learning is an increasingly useful tool for solving problems, and can solve a broader set of problems when put in the hands of a broader set of people. {\em Few-shot} learning is a type of machine learning designed to enable the model to generalize to new classes with very few examples. In this research, we address two audio classification tasks (speaker identification and activity classification) with the Prototypical Network few-shot learning algorithm, and assess performance of various encoder architectures.
Our encoders include recurrent neural networks, as well as one- and two-dimensional convolutional neural networks.  We evaluate our model for speaker identification on the VoxCeleb dataset and ICSI Meeting Corpus, obtaining 5-shot 5-way accuracies of 93.5\% and 54.0\%, respectively.  We also evaluate for activity classification from audio using few-shot subsets of the Kinetics~600 dataset and AudioSet, both drawn from Youtube videos, obtaining 51.5\% and 35.2\% accuracy, respectively.
\end{abstract}

\section{Introduction}
The speech and signal processing communities were among the earliest adopters of deep learning, which is now used extensively there for tasks ranging from speech recognition \cite{chan16las} to speaker and language identification \cite{richardson15deep}, to audio event detection \cite{wang19temporal}. To work well, however, these methods typically require very large amounts of training data, at a great cost.  
In order to mitigate this limitation, researchers have tried well-known strategies, including unsupervised and semi-supervised learning \cite{thomas13semisup}, transfer learning \cite{kunze17transfer}, and data augmentation. 

In recent years, many {\it few-shot learning} methods have been introduced, designed to generalize effectively to unseen {\it classes} with only a handful of examples for each class. 
For example, consider a medical practitioner who wishes to build a cough classifier that, given an audio recording, can classify which type of cough (if any) is present in a recording. Such a classifier could help to automate triage by phone in under-served areas by detecting various cough types (e.g. dry, wet, whooping, barking, croupy) that may be indicative of different diseases. In contrast to classical machine learning techniques, using few-shot learning means the practitioner need only collect a very small ($\sim$5) set of examples of the types of cough of interest, significantly lowering the barrier to creating such a system. As another example, one could train a few-shot classifier to identify linguistic or non-linguistic cues of sexual harassment (e.g. phrases or catcalls), to enable detection of sexual harassment from audio.

One approach to few-shot learning is {\it metric learning}, which involves learning an embedding space to compare classes. 
Notable metric-learning few-shot algorithms include Siamese Networks \cite{koch19siamese}, Matching Networks \cite{vinyals16matching}, Relation Networks \cite{sung17relation}, and Prototypical Networks \cite{snell17proto}. These methods have been primarily developed in the computer vision field, but recent work has begun to address few-shot classification of audio.
Pons et al. \cite{pons18audiofs} experiment with prototypical networks, transfer learning, and the combination thereof in order to improve the performance of audio classifiers provided with small labeled datasets. They evaluate on the UrbanSound8k
\cite{salamon14uk} 
and TUT \cite{mesaros16tut} datasets,
reporting 5-way classification accuracies of up to $\sim74$\% and $\sim68$\%, respectively. Anad et al. \cite{anand19fsspeaker} propose utilizing an autoencoder to learn generalized feature embeddings from class-specific embeddings obtained from a capsule network. Performing exhaustive experiments on VoxCeleb \cite{nagrani17voxceleb} and VCTK
datasets, they obtain 5-way speaker classification accuracies of 91.5\% and 96.5\%, respectively.  

While other work has evaluated the effectiveness of few-shot learning with audio with a limited set of encoders on a limited set of datasets, we conduct a study of audio few-shot classification contrasting five audio encoder architectures and reporting results on four widely used audio datasets. We use the prototypical network few-shot algorithm due to its strong performance on image datasets.

\section{Methods}
\begin{figure}
  \includegraphics[width=\linewidth]{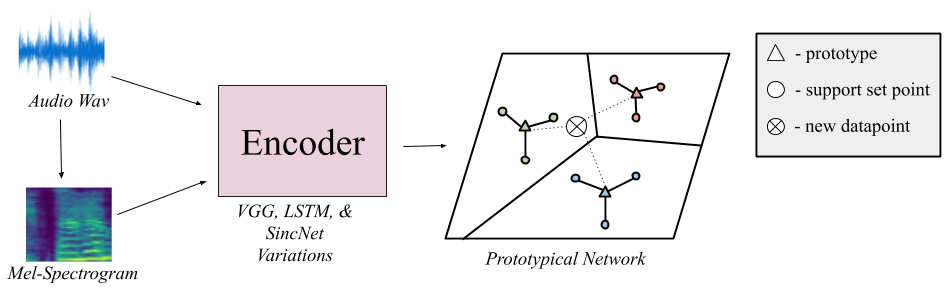}
  \caption{Few-shot pipeline. 3-shot 3-way setup shown. Prototypical Network diagram inspired by a figure in \cite{snell17proto}.}
  \label{fig:fewshot}
\end{figure}

Our pipeline begins with audio input, either raw waveforms or log-scaled mel filterbank features, which are then fed into an encoder, which outputs embeddings that are used by the prototypical network. The pipeline is shown in Fig.~\ref{fig:fewshot}.

\subsection{Feature Extraction} \label{sec:feature_extraction} 
Each datapoint is a waveform audio clip with a 16kHz sampling rate. The SincNet acts directly upon this raw waveform. For the LSTM and VGG11 encoders, we further break the waveform into overlapping 25ms frames with a 10ms offset using a sliding hamming window.  From each frame we extract 64 mel filterbank features (the mel scale is a non-linear frequency scaling aligned with human perception of frequency). The features are then log scaled.

\subsection{Encoders}\label{sec:encoders} 

\subsubsection{VGG11}\label{sec:vgg}
We utilize a VGGish\footnote{https://github.com/tensorflow/models/tree/master/research/audioset/vggish} model, based on the popular VGG
\cite{simonyan14vgg} 2d convolutional neural network (CNN) that has performed well for computer vision tasks. 
Among the model configurations we tried, including deeper variants, VGG11 performed the best.
We feed it windows consisting of 96 consecutive frames, each window offset by 48 frames from the previous.  The 3072-dimensional per-window outputs of VGG11 are averaged to produce the representation of the entire audio clip. 

\subsubsection{LSTM}\label{2}
Long Short-Term Memory networks \cite{hoch97lstm} are a popular type of recurrent neural network; with LSTMs, predictions at timestep $t$ can in theory leverage information contained in all inputs up to and including time $t$. 
We use a single-layer LSTM with a hidden size of 4096 and an output size of 2048. 
At each timestep, we feed in one frame. 
The per-frame outputs are averaged to produce the clip embedding. 

\subsubsection{SincNet}\label{3}
The SincNet \cite{rava18sincnet} is a 1d CNN that acts upon the raw waveform. It is intended to learn meaningful filters in the first layer of the model; specifically, built off of parameterized sinc functions, it learns high and low cutoff frequencies of a set of frequency bins. Prior to training, the model is initialized to the same mel-scale bins used in the mel spectogram.  
We also develop two variants, SincNet+LSTM and SincNet+VGG11, in which the output of the SincNet is fed into LSTM and VGG11 encoders, respectively. 

\subsection{Few-shot Learning and Prototypical Network}\label{sec:fewshot} 
Many few-shot learning methods, including the prototypical network we use here, are trained and evaluated using the concept of an {\it episode} \cite{vinyals16matching}. An episode consists of a support set and a query set. To build an episode, we first randomly draw $k$ classes. The support set consists of $n$ examples from each of these classes, and the query set consists of examples drawn from each of these classes that are not in the support set. This corresponding task is sometimes referred to as a $n$-shot $k$-way classification problem. Episodes function like batches and training proceeds by repeatedly sampling an episode, computing the gradients of the loss function on the query set, and taking a gradient step to adjust model parameters.

The parameters of the overall network are the weights of the encoder that embed each data point into the feature space (the Prototypical Network itself is non-parametric).  In these networks, each class is defined by a prototype, which is simply the average of the embeddings for all support set datapoints belonging to that class. The probability the model assigns to each class for some novel data point $x$ depends on the squared euclidean distance between the embedding of $x$ and the prototypes for each class (see Fig.~\ref{fig:fewshot}).

\section{Experiments}

\subsection{Datasets}\label{BB}

\subsubsection{Kinetics 600}\label{sec:kinetics} 
The Kinetics 600 \cite{carreira18kinetics} dataset consists of 10 second clips of distinct actions (e.g. hugging baby, opening wine bottle). 
We use train/validation/test splits that are suitable for a few-shot setting, proposed by \cite{careaga19video}. There are two test sets defined: one with randomly selected held out classes, and the other with held out musical classes (instruments, singing, etc.) which are more likely to be discriminable by audio.

\subsubsection{VoxCeleb}\label{sec:voxceleb} 
VoxCeleb \cite{chung18voxceleb2} is a dataset containing hundreds of thousands of utterances of celebrity speech. These audio clips are recorded in a diverse set of acoustic environments, ranging from outdoor stadiums to indoor studios, with varying quality. We create splits suitable for the few-shot setting, with speakers as classes.

\subsubsection{ICSI Meetings Corpus}\label{sec:icsimc} 
The ICSI Meetings Corpus \cite{janin03icsi} consists of natural meetings held at the International Computer Science Institute in Berkeley, California. In order to utilize this data for the few-shot setting, we use a subset of the corpus containing meetings that only have speakers that are also present in other audio clips (so that we can build a support set for each query speaker). In total, we use 64 of the meetings, and segment them using ground truth segmentations to produce the datapoints used in the support and query sets. Due to the limited amount of data, we pretrain our VGG11 model on VoxCeleb and evaluate on ICSI without further training.

\subsubsection{AudioSet}\label{4} 
AudioSet \cite{gemmeke17audioset} is a collection of audio from YouTube videos, specifically chosen for acoustic content. 
Because there are multiple positive labels per audio clip, we create a subset of AudioSet suitable for a few-shot setting. With a discrete optimization algorithm, we find an approximately optimal subset of classes that maximizes the number of audio clips containing only a single positive label among the subset of classes chosen.  
This yields a set of 150 classes, each having at least 378 examples, which is then split into train, validation and test.

\subsection{Training Details}\label{CC}
All experiments have $n=5$ labeled datapoints per class in the support set, with $k=1$ or $k=5$ classes in each episode. We perform additional experiments with $k=10$ for AudioSet. 
We train for 25,000 episodes, but evaluate validation performance every 500 episodes and use early stopping to terminate training if no progress has been made on the validation set for 10 consecutive checks.
We use the Adam optimizer with a fixed learning rate of $10^{-5}$. Little to no hyper-parameter tuning is performed. After training converges, we evaluate on 1000 randomly selected episodes from the test set and report average performance across these episodes.

\subsection{Results and Discussion}
We evaluate  each of the encoders on Kinetics 600 and VoxCeleb and report the results in Tables~\ref{tab:kinetics} and \ref{tab:voxceleb}.
The results show that VGG11 is the best performing model across both tasks (activity classification and speaker identification). None of the SincNet variants outperforms VGG11 alone, but SincNet+LSTM does outperform the LSTM and SincNet individually. Note that the models perform better on our Kinetics 600 Test Set 2, consisting of different musical instruments, than Test Set 1, as would be expected of a set whose classes are defined by acoustic cues. 
The speaker identification task (Table~\ref{tab:voxceleb}) is significantly easier than the activity classification task, with very strong performance on VoxCeleb.

Given that VGG11 yields the best performance, we evaluate it on the ICSI Meetings Corpus. Results are reported in Table~\ref{tab:icsi}, which lists
speaker identification accuracies as a function of the number of speakers in the file (averaging across files with the same number of speakers).
As expected, accuracy generally gets better as the number of speakers decreases. 

Finally, our results on AudioSet are shown in Table~\ref{tab:audioset}.  
Performance is very similar to Kinetics. On one hand, this is expected as both AudioSet and Kinetics are derived from Youtube, and thus highly noisy and variable.  On the other hand, this is somewhat surprising because the AudioSet clips were specifically chosen to be characterized by audio, and therefore we would anticipate better performance on it.  

\begin{table}
\begin{center}
 \begin{tabular}{||c c c||} 
 \hline
  & 1-shot, 5-way & 5-shot, 5-way \\ 
 \hline\hline
 VGG11 & {\bf 35.1\%} / {\bf 35.3\%} & {\bf 47.8\%} / \textbf{51.5\%} \\ 
 \hline
 LSTM & 27.7\% / 30.2\% & 38.1\% / 42.2\% \\
 \hline
 SincNet & 27.5\% / 31.4\% & 34.7\% / 41.5\% \\
 \hline
 SincNet+VGG11 & 31.2\% / 34.9\% & 44.9\% / 48.4\% \\
 \hline
 SincNet+LSTM & 29.5\% / 30.6\% & 38.5\% / 46.3\% \\ 
 \hline
\end{tabular}
\end{center}
\caption{Kinetics 600 Test Set 1 / Test Set 2 Accuracies.} \label{tab:kinetics}
\end{table}

\begin{table}
\begin{center}
 \begin{tabular}{||c c c||} 
 \hline
   & 1-shot, 5-way & 5-shot, 5-way \\ 
 \hline\hline
 VGG11 & {\bf 79.9\%} & \textbf{93.5\%} \\ 
 \hline
 LSTM & 68.4\% & 86.5\% \\
 \hline
 SincNet & 37.5\% & 49.9\% \\
 \hline
 SincNet+VGG11 & 64.8\% & 82.1\% \\
 \hline
 SincNet+LSTM & 70.5\% & 88.3\% \\ 
 \hline
\end{tabular}
\end{center}
\caption{VoxCeleb Test Set Accuracies} \label{tab:voxceleb}
\end{table}

\begin{table}
    \begin{center}
    \begin{tabular}{||ccccccc||}\hline
    \# Speakers & 4 & 5 & 6 & 7 & 8 & 9\\\hline
    Avg. Acc.& 66.4\% & 54.0\% & 56.5\% & 53.1\% & 59.3\% & 36.9\%\\\hline
    \end{tabular}
    \end{center}    
    \caption{ICSI Meetings Corpus Accuracies with VGG11} \label{tab:icsi}
\end{table}

\begin{table}
\begin{center}
\begin{tabular}{||c c c c||} 
\hline
 1-shot 5-way & 5-shot 5-way & 1-shot 10-way & 5-shot 10-way \\ 
\hline\hline
31.0\% & 35.2\% & 19.3\% & 25.0\% \\ 
\hline
\end{tabular}
\end{center}
\caption{AudioSet Test Set Accuracies with VGG11} \label{tab:audioset}
\end{table}

\section{Conclusions}
Few-shot learning methods aid efforts to democratize machine learning, giving people the ability to construct classifiers that solve problems that matter to them, with fewer resource hurdles. This should boost machine learning's applicability to a long tail of tasks with societal impact. 
In this paper, we provide a study of few-shot learning applied to audio data. Our experiments cover four different datasets, split between speaker identification and activity classification tasks. 
We compare the performance of three existing audio encoder models, and propose two new variations (SincNet+VGG11 and SincNet+LSTM). 
We find that the VGG-based model performs the best across all datasets and tasks. 
Useful extensions of this work left to future work include varying the few-shot method itself, and applying these findings to problems that directly impact underrepresented groups in technology.

\balance
\bibliography{main}
\bibliographystyle{unsrt} 

\end{document}